# Local Electrostatic Imaging of Striped Domain Order in LaAlO$_3$/SrTiO$_3$

M. Honig*[1], J. A. Sulpizio*[1], J. Drori[1], A. Joshua[1], E. Zeldov[1], and S. Ilani[1]

[1]*Department of Condensed Matter Physics, Weizmann Institute of Science, Rehovot 76100, Israel.*

\* These authors contributed equally to this work

**The emerging field of complex oxide interfaces is generically built on one of the most celebrated substrates—strontium titanate (SrTiO$_3$). This material hosts a range of phenomena, including ferroelasticity, incipient ferroelectricity, and most puzzlingly, contested giant piezoelectricity. Although these properties may markedly influence the oxide interfaces, especially on microscopic length scales, the lack of local probes capable of studying such buried systems has left their effects largely unexplored. Here we use a scanning charge detector—a nanotube single-electron transistor—to non-invasively image the electrostatic landscape and local mechanical response in the prototypical LaAlO$_3$/SrTiO$_3$ system with unprecedented sensitivity. Our measurements reveal that on microscopic scales SrTiO$_3$ exhibits large anomalous piezoelectricity with curious spatial dependence. Through electrostatic imaging we unravel the microscopic origin for this extrinsic piezoelectricity, demonstrating its direct, quantitative connection to the motion of locally ordered tetragonal domains under applied gate voltage. These domains create striped potential modulations that can markedly influence the two-dimensional electron system at the conducting interface. Our results have broad implications to all complex oxide interfaces built on SrTiO$_3$ and demonstrate the importance of microscopic structure to the physics of electrons at the LaAlO$_3$/SrTiO$_3$ interface.**

Correspondence to: joseph.sulpizio@weizmann.ac.il

Until now, the main thrust in the study of the LaAlO$_3$/SrTiO$_3$ interface[1,2] (LAO/STO) has been the exploration of its conducting two-dimensional electron gas (2DEG), which exhibits unique properties such as superconductivity[3,4], ferromagnetism[5], and their coexistence[6-9]. Much less explored is how this interface is affected by the diverse array of physical phenomena of its hosting STO material. At $T = 105 K$ STO undergoes a ferroelastic transition from cubic to tetragonal crystal symmetry[10], and at $40\ K$ it becomes a quantum paraelectric[11]. It has even been suggested that at much lower temperatures ($T < 10\ K$) this material could have a diverging piezoelectric response[12], much larger than that of conventional piezoelectrics. Whereas for ferroelectric materials it has been clearly demonstrated[13,14] that such anomalous piezoelectricity can arise from the motion of structural domains, the origin of the unexpectedly large cryogenic piezoelectricity in a centrosymmetric material such as STO has remained unresolved. These structural aspects could also have a microscopic imprint on the physics of the electrons at the LAO/STO interface, prompting the need for imaging its physics on the nanoscale. Whereas scanning superconducting quantum interference device measurements[9] have provided crucial information on its magnetic properties, progress towards microscopic imaging of its electrostatic and mechanical properties has remained elusive.

The challenge in imaging the conducting interface is that it is buried underneath the surface. This makes the use of classical surface probes such as STM, which rely on tunnelling, quite demanding[15]. Approaches based on atomic force microscopy can accurately characterize its mechanical properties (piezo force microscopy[16,17]); however, their use as electrostatic detectors (Kelvin probes)[18] is quite limited, because they do not have sufficient energy resolution for probing the highly screened potential features in the 2DEG below the surface. A less conventional detector that provides substantial improvement in energy resolution is the scanning single-electron transistor[19] (SET), which proved instrumental in studying the microscopic physics in semiconductor heterointerfaces[20].

In this work we introduce a new generation of scanning SETs (ref. 21), based on carbon nanotubes, capable of imaging on the nanoscale both mechanical properties and electrostatic landscapes with an unprecedented energy resolution. We use this SET to uncover the importance of ferroelastic domains to the anomalous piezoelectricity in STO and their effect on the electrons at the LAO/STO interface.

Our scanning SET, shown schematically in Fig 1a, is comprised of a short semiconducting nanotube, connected at the end of a scanning probe cantilever to source and drain electrodes and suspended above a gate (Supplementary Section 1). Using this local gate we form an electronic quantum dot in the suspended section of the nanotube (red), separated by a pair of p-n junction barriers from the hole-doped sections near the contacts (blue). At cryogenic temperatures, the current through the dot exhibits sharp Coulomb blockade oscillations as a function of the induced charge $Q = C\phi$, where $\phi$ and $C$ are the electrostatic potential difference and capacitance, respectively, between the nanotube and the sample under study (Fig. 1b). Monitoring this current, we can resolve a tiny fraction of a single-electron charge: $\delta Q \sim 2 \times 10^{-5} eHz^{-1/2}$.

The scanning SET images various physical quantities through the different ways in which the sample induces charge on the dot: $\delta Q = C\delta\phi + \phi\delta C$. The first term is proportional to $\delta\phi$ and reflects changes in the potential of the sample, whereas the second is proportional to $\delta C$ and reflects changes in the vertical displacement of the sample surface. As we separately control the prefactors of these terms, we can independently measure both electrostatic and mechanical properties of the LAO/STO system (details in Supplementary Section 2). Consider the simplest example of a sample containing two homogeneous domains, separated by a domain wall, whose surface potentials differ by $\Delta\phi$ (Fig. 1c). By monitoring the SET current while scanning it across this wall, we can detect a change in the gating of the nanotube by $\Delta\phi$ and thus map the surface potential as a function of position. Further, if the wall moves laterally with the application of an oscillating back-gate voltage ($V_{BG}$) with respect to the 2DEG, $\delta V_{BG}$, its associated potential step, $\Delta\phi$, will also oscillate laterally. By detecting the local potential response with respect to this excitation, $d\phi/dV_{BG}$, which we term the lateral electromechanical response, we can directly observe laterally moving domain walls (Fig. 1d). Finally, we can measure the local vertical displacement of the sample surface in response to $\delta V_{BG}$ through the corresponding change in the local nanotube-sample capacitance, $dC/dV_{BG}$. By normalizing this capacitance change with the capacitance change obtained by oscillating the SET-sample separation with a given amplitude, $dC/dz$, we directly determine the vertical electromechanical response, $dz/dV_{BG}$. This response is the local piezoelectric coefficient $d_{333}$ (Fig. 1e). Measurements of $C$ also enable imaging of the surface topography. We emphasize that in contrast to STM, where current tunnels from the tip to the sample, and to piezo force microscopy, which requires mechanical contact between the tip and sample, the scanning SET involves neither charge transfer nor mechanical contact with the sample, and thus acts as a non-invasive local detector.

We begin by measuring the local piezoelectric response, parking the SET at a fixed lateral position above the LAO/STO sample, and probing $dz/dV_{BG}$ as a function of $V_{BG}$. The sign of the measured piezoelectric response (Fig. 2a) is gate voltage dependent, and its magnitude is anomalously large (up to $1 nmV^{-1}$), similar to the values reported by macroscopic measurements at our measurement temperature of $T = 4K$ (ref. 12; although here we measure a different element of the piezoelectric tensor, $d_{333}$ rather than $d_{311}$). Over the 400V gate voltage span, the surface moves hundreds of nanometers (showing some hysteresis, Supplementary Section 4), establishing that the response arises within the STO. Curiously, we also find that the piezoelectric coefficient changes quite sharply near two gate voltages (dashed lines in Fig 2a). These transitions divide the curve into three regions with clearly different piezoelectric response within the STO.

To check whether the abrupt change in piezoelectric response occurs homogeneously over the entire sample or has an associated spatial structure, we image the local piezoelectricity over a $30 \mu m \times 30 \mu m$ window at various gate voltages around the sharp transition at negative $V_{BG}$. On both sides of the transition, the response is spatially homogeneous (Fig. 2b,d). However, at the transition (Fig. 2c) a clear boundary emerges, separating distinct piezoelectric regions in real space. Interestingly, measurement of the surface topography reveals that it is kinked at the boundary (Fig. 2e) with a small but discernible angle, $tan(\alpha) \approx 1/1000$.

Having characterized the local vertical mechanical response, we proceed to measure the lateral electromechanical response, $d\phi/dV_{BG}$ (Fig. 1d), which allows us to extract the gate-dependent physics from the large fixed disorder potential background ($\sim 10 - 50\ mV$, Supplementary Section 5). Figure 3a shows a map of $d\phi/dV_{BG}$ measured over the transition region corresponding to Fig 2c. This map reveals an intricate pattern of stripes with remarkable correlation to the piezoelectric response: the lower piezoelectricity region is homogeneous in the lateral electromechanical response, whereas the upper piezoelectricity region is highly striped. Interestingly, nearly all stripes come in pairs of positive (red) and negative (blue) signs, which are oriented either vertically or horizontally. A notable exception is a set of unpaired (bottom-most) horizontal blue stripes at the boundary between regions of different piezoelectricity. Importantly, the magnitude of the response is identical within each pair of stripes, showing that they correspond to upward and downward potential steps of equal magnitude displaced laterally by equal amounts by the a.c. gate voltage. The origin of these stripes is uncovered through optical imaging experiments performed on similar LAO/STO samples (Supplementary Sections 6-9). These experiments show combs of stripes, just like those measured by the SET, and find that these stripes appear sharply below $T = 105\ K$, establishing that they arise from the ferroelastic transition of STO. The observed stripes are therefore the walls separating domains with different tetragonal orientations.

We reconstruct the full map of domain orientations from the electromechanical images by using the simple tiling rules of tetragonal domains. To minimize dislocations, tetragonal unit cells along the $X$ (100), $Y$ (010) or $Z$ (001) crystal directions (Fig. 3b) tile such that they share their $a$ axis (Fig. 3c). Consequently, at the top surface, the boundary between $X$ and $Y$ domains must lie at 45° or 135°, between $Z$ and $X$ at 0°, and between $Z$ and $Y$ at 90° (Fig. 3c). The vertical (horizontal) stripes in the measurement thus correspond to boundaries between $Z$ and $X$ ($Z$ and $Y$) domains. The top half of Fig. 3a therefore consists of alternating sets of $Z$ and $X$ domains (vertical) and $Z$ and $Y$ domains (horizontal; illustrated in Fig. 3d). The unpaired blue stripes separating the piezoelectric regions are then boundaries to a large $Y$ mono-domain filling the bottom of the image.

A crucial observation relating the observed domains to the anomalous STO piezoelectricity is that they move readily with varying gate voltage. Figure 4a shows a series of lateral electromechanical response maps taken at increasingly negative gate voltages through the transition at negative $V_{BG}$ in Fig. 2a. Tellingly, the boundary between the striped and homogeneous regions smoothly sweeps across the field of view at a typical rate of $1\ \mu m V^{-1}$, until at sufficiently negative gate voltages $d\phi/dV_{BG}$ is entirely homogeneous. Domains within the striped region also slide to the right with decreasing gate voltage at similar rates, although their dynamics show a richer structure of discrete steps (Fig. 4b). Optical measurements provide a larger field of view (Fig. 4c) and show that with increasingly negative $V_{BG}$ the striped regions retreat from the bulk of the sample to its edge (Supplementary Sections 6 and 8). Identical transitions between striped and homogeneous regions are seen in several samples, including bare STO, suggesting that this phenomenon is quite general and robust. The motion of the domains under application of gate voltage may arise from dielectric constant and elastic moduli differences between in-plane and out-of-plane domains[22], leading to anisotropic electrostriction[23]. It may also result from the domain walls being charged or polar, as suggested by previous work[24-26], providing a natural coupling to an applied gate voltage.

Combining the above measurements, we can now understand the origin of anomalous piezoelectricity in STO. Three observations should be taken into account: the two regions with distinct piezoelectric response have tetragonal domains oriented within the plane (homogeneous region) and predominantly normal to the plane (striped region); their separating domain wall moves with a typical rate of $\sim 1\ \mu m V^{-1}$; and the surface is kinked at this wall with a change in slope of $tan(\alpha) \approx 1/1000$. This kink translates the in-plane motion of the domain wall to motion of the surface perpendicular to the plane with a speed of $1\ nm V^{-1}$ (Fig. 4d), in excellent agreement with the giant piezoelectricity that we measure. In fact, a kinked surface between $Z$ and $Y$ ($Z$ and $X$) tetragonal domains is expected, because it is required for matching the different extension of their unit cells in the $z$ direction. As seen in the illustration in Fig. 4d, the angle across the kink is simply $tan(\alpha) = \frac{c}{a} - 1 \approx 1/1000$, where we use the known[27] bulk value of $c/a$ at $T = 4K$. This is in excellent agreement with our measurements presented in Fig. 2e. The measurements above thus reveal a direct and quantitative microscopic origin for the giant extrinsic piezoelectricity of STO - the motion of tetragonal domains.

The emergence of tetragonal domains leads also to observable potential stripes with further important implications for the 2DEG at the interface. The effect becomes clear when we measure directly the potential of the surface above the various domains. Figure 5a,b shows the imaged potential at the domain surfaces with the corresponding lateral electromechanical signal (details in caption). Notably, we see that in-plane tetragonal domains ($X$ and $Y$) have a similar surface potential, which is different by $\sim 1\ mV$ from that of the out-of-plane domains ($Z$). This is most clearly seen when plotting the potential along a line cut across the map (Fig. 5c), showing that after averaging the remnant disorder the potential is roughly constant within each type of domain, but varies digitally by $\sim 1\ mV$ between the $X$ and $Z$ domains. This difference in the surface potential between domains requires charge to be transferred across the domain wall, which for typical parameters (Supplementary Section 11) would lead to a significant modulation of the local 2DEG density, up to tens of per cent of the average density.

Such a possibly large modulation of the 2DEG density across domain walls can substantially impact the macroscopic transport phenomena measured at the interface (for example, refs 28-30). As the 2DEG mobility depends strongly on its density[31], such a modulation may lead to 2DEG mobility that strongly varies in space and a corresponding channeled flow between domains or at the domain walls[32]. Indeed, recent complementary measurements on LAO/STO using scanning magnetometry in the presence of transport current show such strongly channeled flow within the 2DEG (ref. 33), closely matching our measured striped potential. In light of these observations, previous and future transport and other macroscopic measurements should be carefully examined for the influence of domain physics. We note however that the spread of domains strongly depends on the cycling of $V_{BG}$ (Supplementary Sections 8 and 9). In all samples we have measured, we find that domain walls retreat irreversibly to the edges once the back gate is cycled to high voltages ($200-400\ V$ in magnitude), with the bulk remaining free of $Z$ domains[23]. We do observe, however, that the domain structure may remain pinned within the bulk of the sample by evaporated metal, wirebonds and possibly other lithographic features. All of these factors should be taken into account when designing and interpreting LAO/STO transport or other macroscopic experiments.

Using a new scanning SET probe, we have demonstrated that tetragonal domains in STO give rise to its anomalous extrinsic piezoelectricity, as well as generate striped potential modulations that can markedly influence the 2DEG at the LAO/STO interface. These results are important for a wide variety of complex oxide interfaces that are generically built on STO. While this work suggests that the tetragonal domains should be kept in mind when designing and interpreting experiments, it also offers surprising new avenues for harvesting[17] the domain physics of STO through strain engineering[34] and the opportunity to study a unique class of low dimensional structures formed within a correlated, degenerate electron system.

**Methods**

Sample fabrication: LAO/STO samples were grown via pulsed laser deposition of LaAlO$_3$ on single crystals of TiO$_3$-terminated (001) SrTiO$_3$. Growth was carried out at $\sim 10^{-4}$ mbar of O$_2$ with a laser repetition rate of $1\ Hz$ and pulse fluence of $0.6\ Jcm^{-2}$ at $T = 650°C$. The samples were thermally annealed after growth for ~1 hour at $600°C$ in 200 mbar of O$_2$. SET measurements were performed on unpatterned 6uc LAO/STO (2DEG characterization in Supplementary Section 10) approximately $400\mu m$ from the sample edge, and optical measurements performed on 10uc LAO/STO as well as bare STO. The STO thickness used throughout is $500\ \mu m$.

Measurements: The back gate voltage of the samples was cycled ($\pm 200V$ for SET measurements, $\pm 400V$ for optical measurements) to obtain repeatable tetragonal domain distributions. Scanning SET measurements were performed in a home-built scanning probe microscope setup at a temperature of $T = 4K$. The SET spatial imaging resolution in the measurements reported in this paper is $\sim 600nm$ (Supplementary Section 3). Various physical quantities were measured simultaneously using lock-in based techniques (Supplementary Section 2). The measurements were verified to be free of $RC$ time constant effects over the frequency span used $(2 - 300 Hz)$. Optical measurements were performed in a flow cryostat under a polarized light microscope.

**Acknowledgements**

We acknowledge S. Gariglio and J.-M. Triscone for providing the samples and for useful discussions and comments on the manuscript. We thank Y. Yacoby, B. Kalisky, E. Berg, E. Altman, P. Paruch, J. Ruhman, P. Zubko, L. Yu and A. Zunger for helpful discussions. We note that a previous experiment using a complementary scanning superconducting quantum interference device technique[33] has observed enhanced conductivity due to tetragonal domain structure at this oxide interface. S.I. acknowledges financial support by the Israel Science Foundation (No. 1267/12), the Minerva Foundation, the ERC Starting Grant (QUANT-DES-CNT, No. 258753), and the Marie Curie People Grant (IRG, No. 239322).

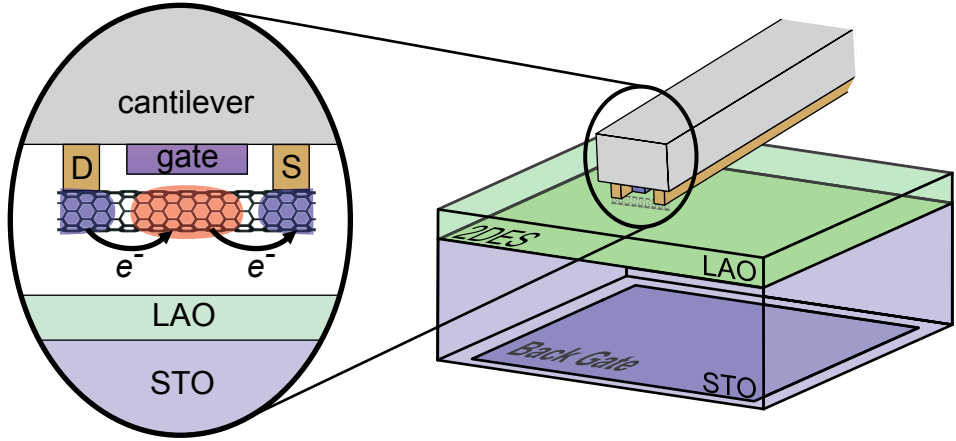

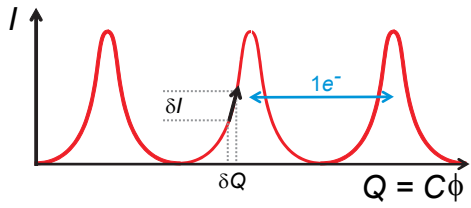

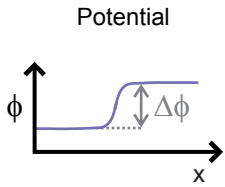

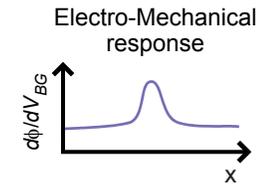

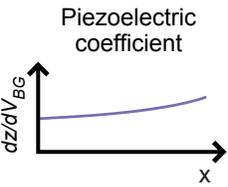

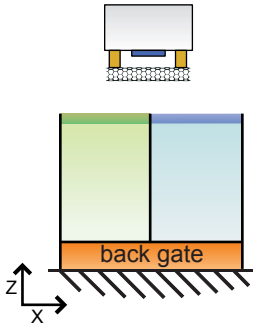

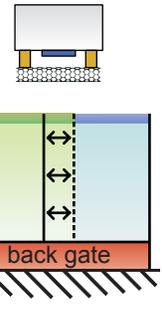

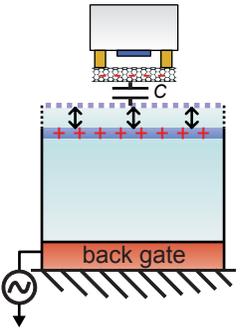

figure 1

**Figure 1. Nanotube-based SET imaging. a**, Illustration of the measurement set-up. A nanotube SET probe is scanned above the surface of a back-gated LAO/STO sample. The nanotube is placed at the end of a cantilever (zoom-in inset), connected to source (S) and drain (D) contact electrodes (yellow) separated by $800\,nm$, and suspended over a gate electrode (purple). At low temperatures an electron quantum dot is formed in the suspended segment (red), separated by a pair of insulating p–n junctions from hole-doped segments near the contacts (blue). **b**, The SET current, $I$ (shown schematically), exhibits Coulomb blockade oscillations as a function of the charge induced on the dot, $Q = C\phi$, where $\phi$ and $C$ are the electrostatic potential difference and capacitance between the sample surface and the nanotube, respectively. Potential and capacitance changes lead to separable contributions to the induced charge, allowing independent measurement of electronic and mechanical effects. **c**, The local electrostatic potential, $\phi$, is imaged by monitoring the SET current while scanning its position along the surface. A specific example is illustrated where the surface potential changes in a step-like manner above a domain wall within the sample. **d**, The lateral electromechanical response, $d\phi/dV_{BG}$, is imaged by oscillating $V_{BG}$ with respect to the 2DEG and measuring the resulting potential oscillations. A domain wall moving with gate voltage will produce a local response in this quantity. **e**, The local piezoelectric response (vertical electromechanical response), $dz/dV_{BG}$, is imaged by monitoring the oscillating nanotube–sample capacitance resulting from an oscillating $V_{BG}$, $\delta C/\delta V_{BG}$, and normalizing by the capacitance change induced by oscillating the nanotube–sample separation by a given amount, $\delta C/\delta z$. Measurements of $C$ as a function of the lateral position of the SET are further used to image the surface topography.

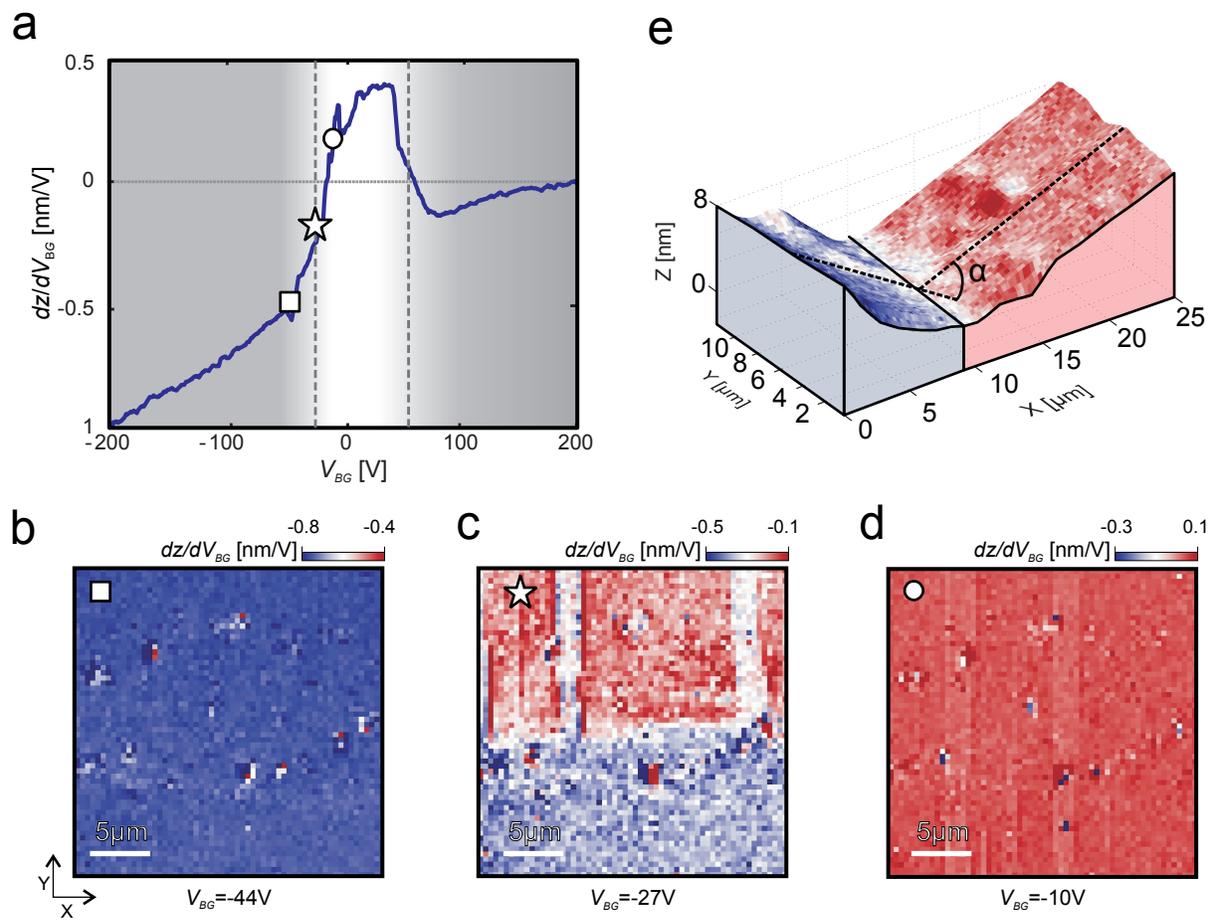

figure 2

**Figure 2. Measurements of the local piezoelectric response. a**, Piezoelectric response, $dz/dV_{BG}$, measured at a fixed point at the LAO/STO surface as a function of $V_{BG}$. Anomalously large $dz/dV_{BG} \sim 1 nmV^{-1}$ is observed, exhibiting three distinct regimes separated by sharp transitions (dashed vertical lines). **b–d**, Spatial maps of $dz/dV_{BG}$ measured at three fixed $V_{BG}$ values (values indicated, corresponding symbols are shown in **a**; an identical range is spanned by the color maps in all frames). Apart from sparse features due to localized disorder, the maps in **b** and **d** seem spatially homogeneous. The map in **c** corresponds to the sharp transition in **a** and is divided into two regions with distinct values of $dz/dV_{BG}$. The value of the response in the lower half of the image is also seen to bleed into several narrow, vertical regions in the upper half of the image. Weak stripe-like features are also faintly visible in **d**. **e**, Measured surface topography corresponding to the mixed regime in **c**, revealing a surface kink with a change in slope of $tan(\alpha) \sim 1/1000$. The surface is colored with the measured local piezoelectric response to demonstrate the strong correlation between the sharp transition in piezoresponse and the topography.

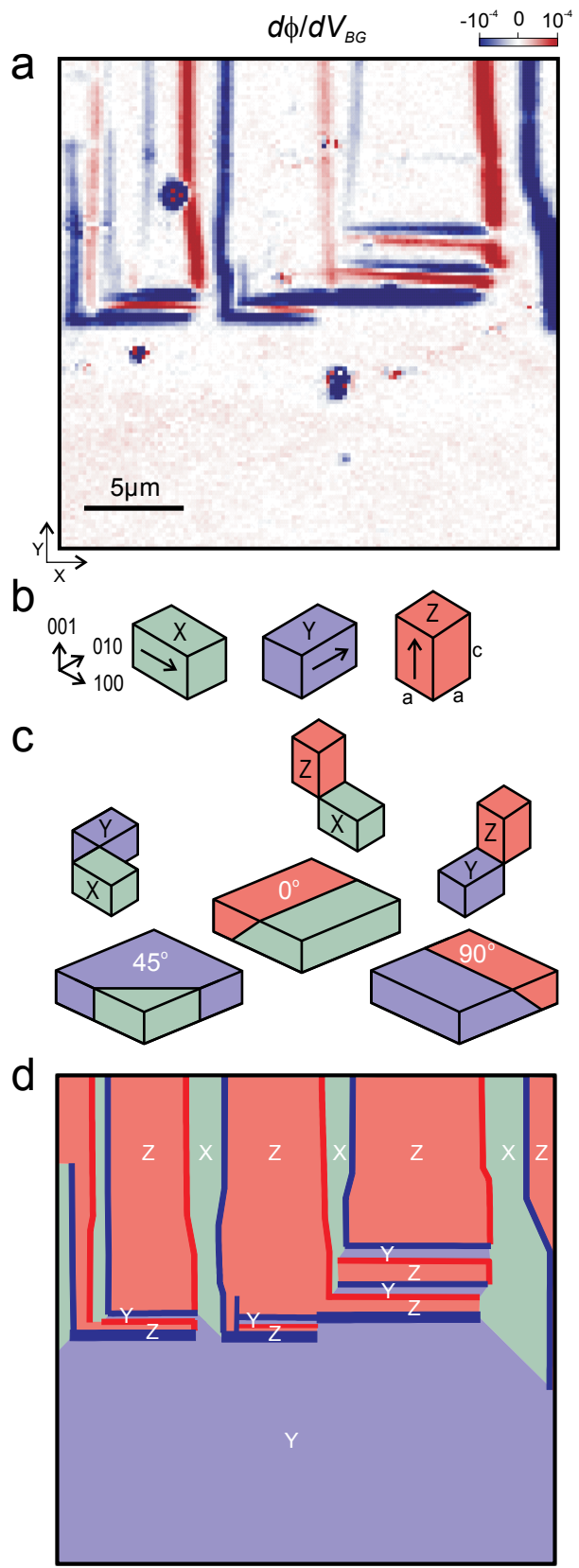

figure 3

**Figure 3. Observation of tetragonal domains within STO. a**, Map of the lateral electromechanical response, $d\phi/dV_{BG}$, around the sharp transition at $V_{BG} = -27V$. The bulk of the image is colored white, indicating negligible lateral electromechanical response, whereas the upper half of the image contains a network of paired red and blue stripes elongated in both the $x$- and $y$-directions. The stripes are peaks of alternate response, each representing the motion of either the rising (red) or falling (blue) edge of a potential step at the surface. A series of unpaired blue horizontal stripes appears along the horizontal boundary observed in Fig. 2c. **b**, Tiling rules of tetragonal domains in STO: the three possible domains are labelled *X,Y,* and *Z* according to the orientation of their long axis ($c$ axis). **c**, To minimize dislocations, intersections of tetragonal domains of different orientations must share their short axis ($a$ axis), forming twin boundaries with well-defined angles. **d**, A schematic map labelling the domain orientations in **a**. The red and blue stripes are the domain wall boundaries. The coloring of the domains follows from the tiling rules in **c**.

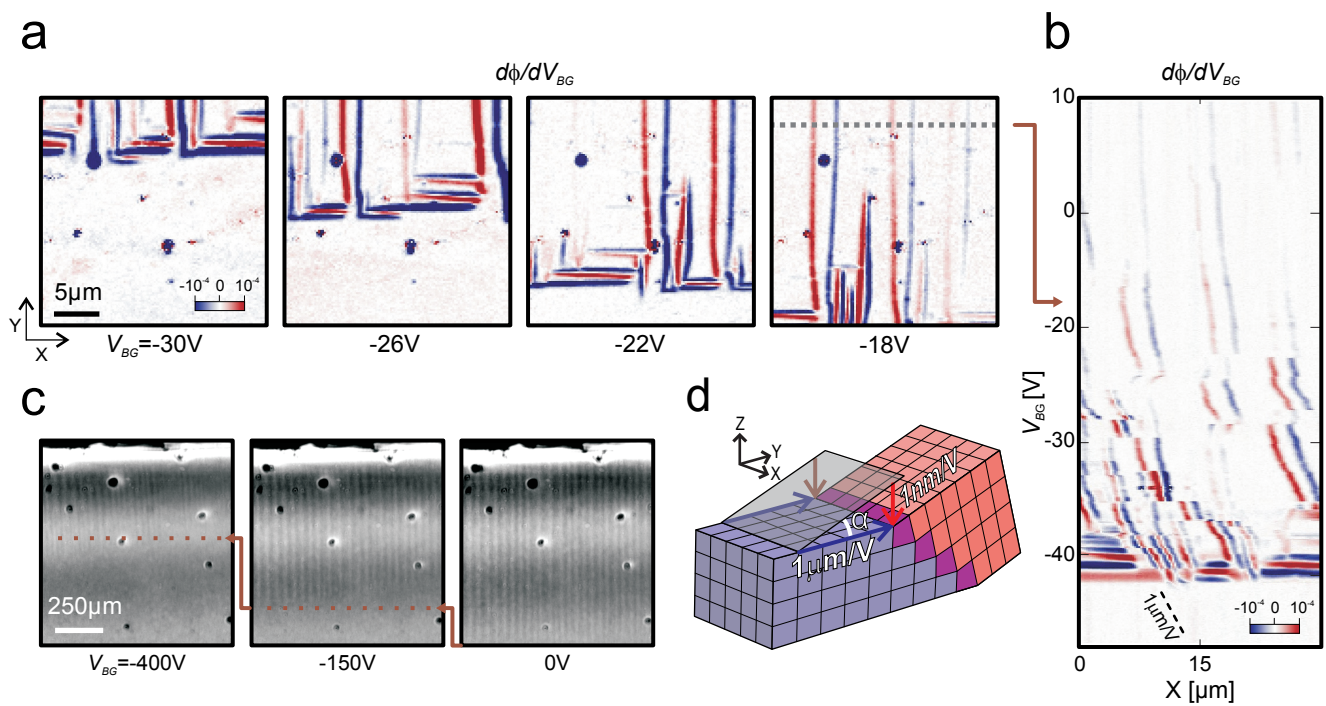

figure 4

**Figure 4. Gate-induced domain motion. a**, Sequence of lateral electromechanical response maps, each taken at a different gate voltage. With decreasing $V_{BG}$ the boundary between the homogeneous and striped regions moves towards the edge of the sample (top) at a rate of $\sim 1 \mu m \, V^{-1}$. **b**, $d\phi/dV_{BG}$ along a horizontal line (grey dotted line in the rightmost frame in **a**), measured as a function of continuously changing $V_{BG}$, demonstrating gate-induced motion of the domains within the striped phase. The domain walls shift to the right, occasionally in discrete jumps, reaching a maximum speed of $\sim 1 \mu m V^{-1}$ (dashed line). Around $V_{BG} \approx -40V$ the piezoelectric boundary crosses the measured horizontal line and the homogeneous response is observed. **c**, Optical images of domain wall motion in a similar LAO/STO sample over a larger field of view, showing that the domains retreat towards the sample edge (top) with a similar rate of $\sim 1 \mu m V^{-1}$ as $V_{BG}$ is made more negative. The brown arrows mark the boundary between the striped domain region and the homogeneous region. **d**, Illustration of the domain wall near the boundary, combining all above results and explaining the origin of the anomalously large piezoelectricity in STO. The surface is angled at the intersection between $Y$ (blue) and $Z$ (red) domains with an angle of $tan(\alpha) = \frac{c}{a} - 1 \approx 1/1000$ (compare with the measured surface topography in Fig. 2e). As the gate voltage is varied the domain wall (purple, penetrating sample at 45°, Supplementary Section 7) moves with a rate of $\sim 1 \mu m V^{-1}$, translating through the kink angle in the surface to a vertical displacement with a characteristic rate of $\sim 1 nm V^{-1}$.

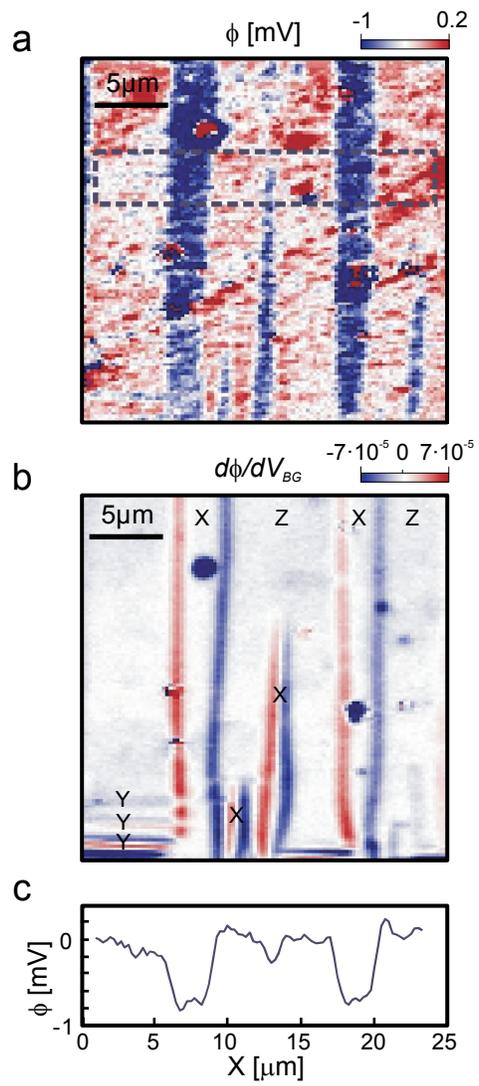

figure 5

**Figure 5. Domain potential and influence on the LAO/STO 2DEG. a**, Measured electrostatic potential map. The domains create a binary potential landscape, where the *Z* domains (red) have one distinct value of potential, and the *X* and *Y* domains (blue) have another value, differing by $\sim 1mV$ ($V_{BG} = -10V$). **b**, Lateral electromechanical response map exhibiting the corresponding domain walls. **c**, Potential profile obtained by averaging along the vertical direction within the dashed rectangle in **a**. The potential steps between *X* and *Z* domains have a characteristic amplitude of $\sim 1mV$.

# SUPPLEMENTARY INFORMATION

## Local Electrostatic Imaging of Striped Domain Order in LaAlO$_3$/SrTiO$_3$

M. Honig*[1], J. A. Sulpizio*[1], J. Drori[1], A. Joshua[1], E. Zeldov[1], and S. Ilani[1]

[1]Department of Condensed Matter Physics, Weizmann Institute of Science, Rehovot 76100, Israel.

* These authors contributed equally to this work

S1. SEM images of the scanning SET probe

S2. Measurement scheme: disentangling mechanical and electronic contributions to the measured response

S3. Scanning SET spatial resolution

S4. Gate voltage dependence of piezoelectric response: asymmetry and hysteresis

S5. Maps of the surface electrostatic disorder potential and the domain-generated surface potential

S6. Optical images of domains: temperature and gate-voltage dependence

S7. Side view optical image of domains spanning entire sample thickness

S8. The dependence of domain distribution on initial voltage cycling post cooldown

S9. Comparison between samples from different suppliers and with different surface treatments

S10. Transport and capacitance characterization of the 2DES of the studied LAO/STO sample

S11. Surface potential and 2DES density modulation

**S1.  SEM images of the scanning SET probe**

A scanning electron microscope (SEM) image of a typical nanotube-based single electron transistor (SET) device is shown in Fig. S1. The SET is situated at the edge of a narrow ($\sim 10 \mu m$) and tall ($\sim 100 \mu m$) cantilever-like pillar, etched in silicon. On this pillar we pattern contact electrodes (yellow, inset Fig S1) and gate electrodes (blue, inset Fig S1) using electron beam lithography. On a separate chip we grow many parallel nanotubes suspended across wide trenches ($\sim 100 \mu m$). We then mate[1] the two chips to place an individual semiconducting nanotube at the pillar's leading edge, as shown in the inset (indicated by black arrows). After mating the nanotube is electrically connected to the source and drain contacts (yellow) and is suspended above multiple gate electrodes (blue). For the current experiments we electrically chain all the gate electrodes together, having them act as a single gate.

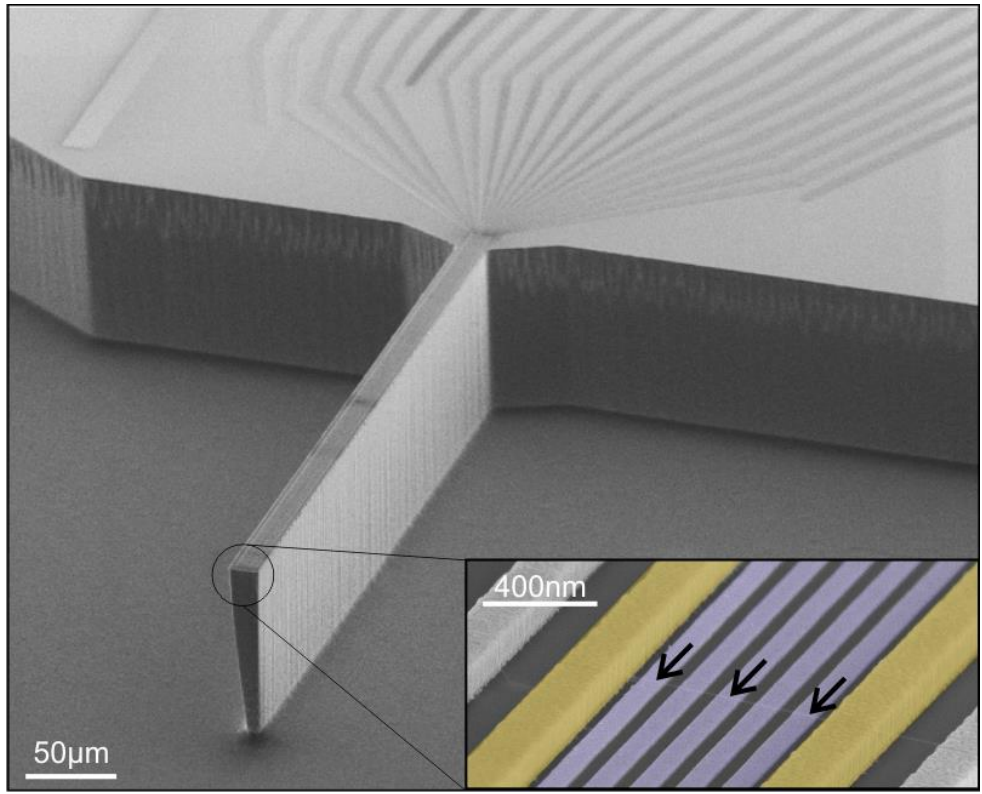

**Figure S1. SEM image of a nanotube-based scanning SET device.** Main panel: A zoom-out view of the deep-etched pillar, where near its leading edge (inset) an individual single-wall carbon nanotube is placed

(indicated by arrows). The nanotube touches source and drain contacts (yellow) and is suspended above five gates (blue). In the experiments in this paper all the gates are electrically chained together acting effectively as one large gate.

## S2. Measurement scheme: disentangling mechanical and electronic contributions to the measured response

The SET senses the local vacuum potential, $\phi$, above the LAO/STO surface, which is related to the energies of the system by (Fig S2a): $\phi = V_S + W$, where $V_S$ is the electrochemical potential, set by a DC voltage source in our experiment, and $W$ is the local workfunction. A change in either of these energies is reflected in a change in $\phi$, which couples to the SET through the induced "control charge", $Q = C\phi$, modulating the measured current as: $\Delta I = g(C\Delta\phi + \phi\Delta C)$, where $g$ is the "gain" of the SET and $C$ is the capacitance between the nanotube and the surface.

The measurement circuit is outlined in Fig. S2b. To independently extract $W$, the vertical piezoresponse, $\frac{\partial Z}{\partial V_{BG}}$, and the lateral electromechanical response embedded in the electronic compressibility, $\frac{\partial \phi}{\partial V_{BG}}$, we exploit our ability to control the electrochemical potential directly with a voltage source, $V_S$. This voltage is applied by a surface contact to the 2DES and the back gate, such that its application would not influence the 2DES charge density, but only change the relative potential of the sample as a whole with respect to the SET. We then simultaneously measure the AC responses in the SET current to several different excitations applied at different frequencies. These include the response to an excitation of the nanotube-surface separation ($H_Z = \frac{\partial I}{\partial Z}$), the electrochemical voltage ($H_S = \frac{\partial I}{\partial V_S}$), and the back gate voltage ($H_{BG} = \frac{\partial I}{\partial V_{BG}}$). To extract the various physical quantities, we form the appropriate ratios of these responses as a function of $V_S$, normalizing out the detailed response characteristics of the SET "amplifier" and directly giving the required physical quantities, as is shown below.

To obtain $W$, we normalize $H_Z$ by $H_S$: $\frac{H_Z}{H_S} = \frac{1}{C}\frac{\partial C}{\partial z} \cdot (W + V_s) = a_Z V_S + b_Z$, which is a function that is linear in $V_S$. The coefficients $a_Z$ and $b_Z$ are obtained by a linear fit in $V_S$. We then find: $W = \frac{b_Z}{a_Z}$.

To obtain the piezoresponse, $\frac{\partial Z}{\partial V_{BG}}$, we normalize $H_{BG}$ by $H_S$: $\frac{H_{BG}}{H_S} = \frac{1}{C}\frac{\partial C}{\partial V_{BG}}(W + V_S) + \frac{\partial \phi}{\partial V_{BG}} = a_{BG} V_S + b_{BG}$, which again is a linear function in $V_S$. Thus, we obtain the coefficients $a_{BG}$ and $b_{BG}$ by a linear fit in $V_S$. Combining these coefficients with those obtained above, we find the piezoresponse: $\frac{\partial Z}{\partial V_{BG}} = \frac{a_{BG}}{a_Z}$. The lateral electromechanical response is then embedded in the extracted electronic compressibility: $\frac{\partial \phi}{\partial V_{BG}} = b_{BG} - a_{BG} \cdot \frac{b_Z}{a_Z}$.

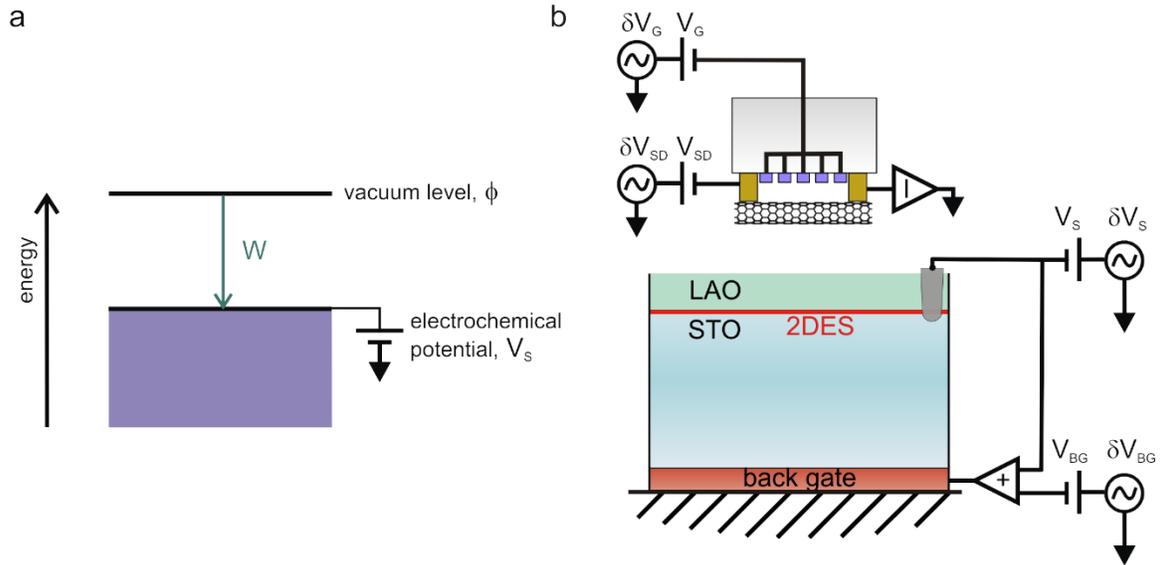

**Figure S2: Scanning single electron transistor measurement scheme.** a) Schematic of energetics at the LAO/STO interface. The vacuum level $\phi$ above the surface is sensed directly by the SET, and is related to the local workfunction $W$ and the electrochemical potential $V_S$ (set directly by an applied voltage in the experiment) by the equation: $\phi = V_S + W$. The shaded region represents the occupied charge states near the surface. b) Measurement circuit: The back-gate, interface, nanotube source contact (yellow, left) and nanotube gates (blue) are connected to DC sources ($V_i$) and AC sources ($\delta V_i$) at different frequencies in the range of 10's to 100's of Hz. The measured current at the NT drain (yellow, right) has AC components at identical frequencies, reflecting its coupling to each of these electrodes. Note that the voltage $V_s$ is connected to both the interface (through an ohmic contact shown in gray) and the back-gate, such that it determines the overall "sample" electrochemical potential. Not shown are three additional AC excitations that oscillate the position of the SET in the X,Y, and Z directions.

### S3. Scanning SET spatial resolution

The spatial imaging resolution of the scanning SET is limited by two terms: the height of the SET above the sample and the dimensions of the electronic quantum dot formed in the suspended nanotube segment, which acts as the actual detector. The size of the quantum dot is set by the distance between the P-N junctions formed between the contact electrodes. These P-N junctions constitute the SET's tunnel barriers, and in general their separation is smaller than the lithographically-defined spacing between the contact electrodes, thus effectively making a smaller detector. The spatial resolution achieved in the measurements presented in this work is $\sim 600 nm$, directly determined by the full width at half maximum of Gaussian fits to peaks in $d\phi/dV_{BG}$ at domain walls, shown in Figure S3.

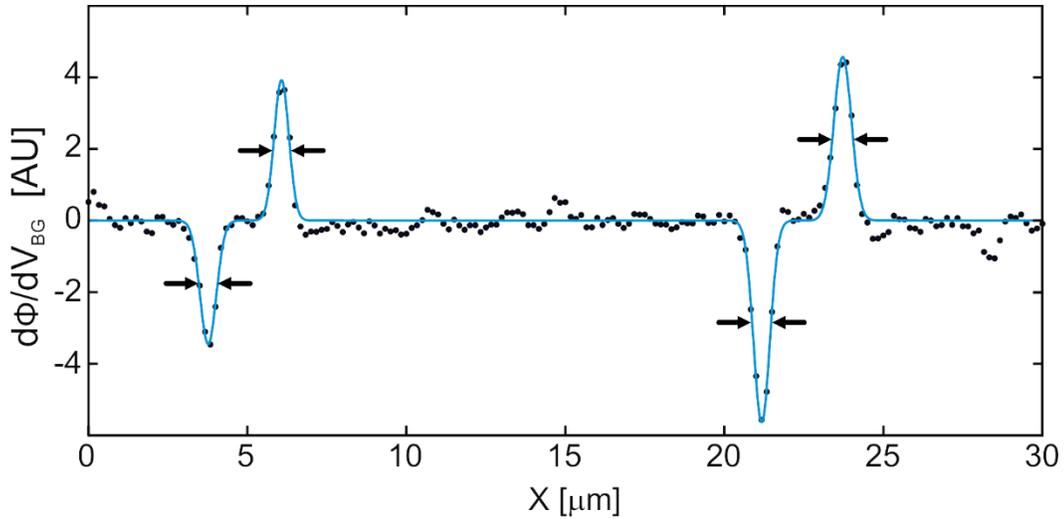

**Figure S3: Scanning SET spatial resolution.** The measured $d\phi/dV_{BG}$ (black dots) is plotted as a function of spatial coordinate across a series of domain walls. The scan is along the direction parallel to the nanotube axis, in which the resolution of the measurement is lower due to the finite length of the quantum dot detector formed in the nanotube along this direction. The solid blue line is a sum of a series of Gaussians fits to the peaks in $d\phi/dV_{BG}$. The full widths at half maximum (FWHM) of the Gaussians (arrows) are $600 nm \pm 40 nm$ giving directly the resolution of our scanning measurements.

**S4. Gate voltage dependence of piezoelectric response: asymmetry and hysteresis**

Figure S4 shows the measured piezoelectric response, $dz/dV_{BG}$, as a function of the back-gate voltage for opposite sweep directions (arrows). The scans highlight two important phenomena related to domain motion: Firstly, $dz/dV_{BG}$ is asymmetric around zero gate voltage – it is large at negative voltages and tends to zero at positive gate voltages. Secondly, around the two sharp transition regions in $dz/dV_{BG}$ this quantity is hysteretic with respect to gate voltage sweep direction. However, the response as measured from voltage sweeps in the same direction is highly repeatable and shows no signs of 2DES depletion over the $400V$ gate voltage span.

The source of the asymmetry and hysteresis becomes clear from optical measurements performed on similar LAO/STO and bare STO samples[2–5] which provide a larger field of view and show domain structure over the entire sample at once. We observe that the asymmetry in piezoresponse is microscopically connected to different regimes of the domain motion at positive and negative gate voltages, as indicated at the top of figure S4. Note that these regimes appear after we have cycled at least once the back gate voltage to large values (see section S8 below), and are measured $\sim 400 \mu m$ from the edge of the sample. At small positive and negative gate voltages we observe striped domain patterns that move strongly with varying gate voltage, which we term the "moving stripes" regime. At large positive gate voltages, domain patterns are still visible, but their motion decreases as the gate voltage is increased. Thus, at these voltages the diminishing piezoelectric response is explained by the absence of domain motion. This is the "fixed stripes" regime. At large negative gate voltages, the stripes retreat to the sample edges as the voltage is made more negative, and a mono-domain fills the image, which we term the "homogenous" regime. In this regime, the scanning field of view is free of domain walls, though beyond this window, the domain walls continue to retreat to the edge of the sample with increasingly negative gate voltage. Thus, the piezoresponse is still large in this regime due to domain motion outside of the field of view of the scanning SET.

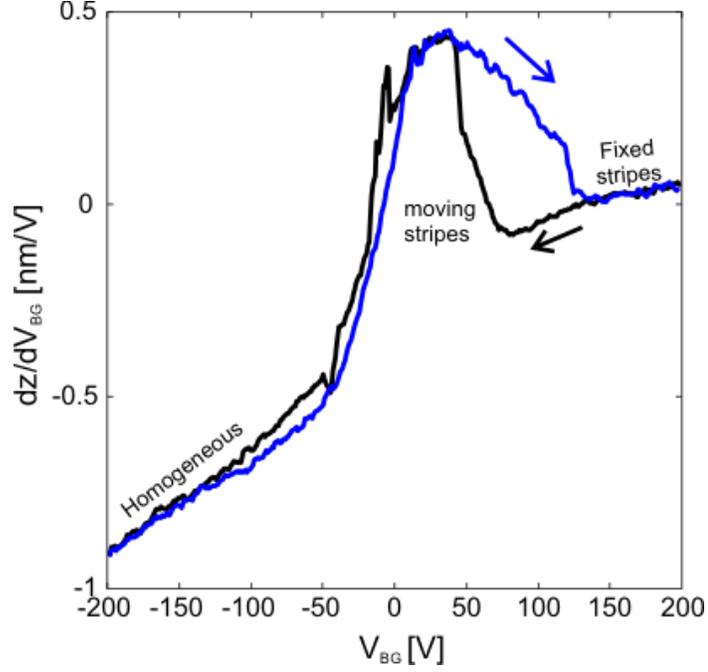

**Figure S4: Piezoelectric hysteresis.** Measured piezoelectric response, $dz/dV_{BG}$, as a function of $V_{BG}$ for opposite sweep directions, indicated by arrows. The three domain regimes are labeled near the corresponding sections of the data.

## S5. Maps of the surface electrostatic disorder potential and the domain-generated surface potential

Figure S5 shows the electrostatic potential measured as a function of position at a fixed back-gate voltage above the LAO/STO surface. The measurement is dominated by surface disorder which is $\sim 10-50 mV$ in amplitude and invariant with respect to the gate voltage. The potential map in (a) is taken at a gate voltage where stripes are present (multiple domains), whereas the potential map in (b) is taken at a gate voltage where the field of view is homogenous (mono-domain). The difference between panels (a) and (b) is shown in (c). By taking differences between these potential maps, we reveal the underlying $mV$-scale potential stripes generated by the tetragonal STO domains. Compare directly with the potential map shown in Fig 5a in the main text.

We note that striped potential modulations can have a more significant impact on the 2DES than the disorder, despite the fact measured disorder is larger in magnitude.

This is due to the fact that the measured disorder potential reflects disorder above the 2DES at the interface, arising from surface absorbents and disorder within the LAO. This bare surface potential will then be substantially screened at the LAO/STO interface by the 2DES. In contrast, the measured striped potential is a measurement of the screened potential modulation at the LAO/STO interface. Thus, from the perspective of the 2DES, the striped potential modulations can be much larger in magnitude.

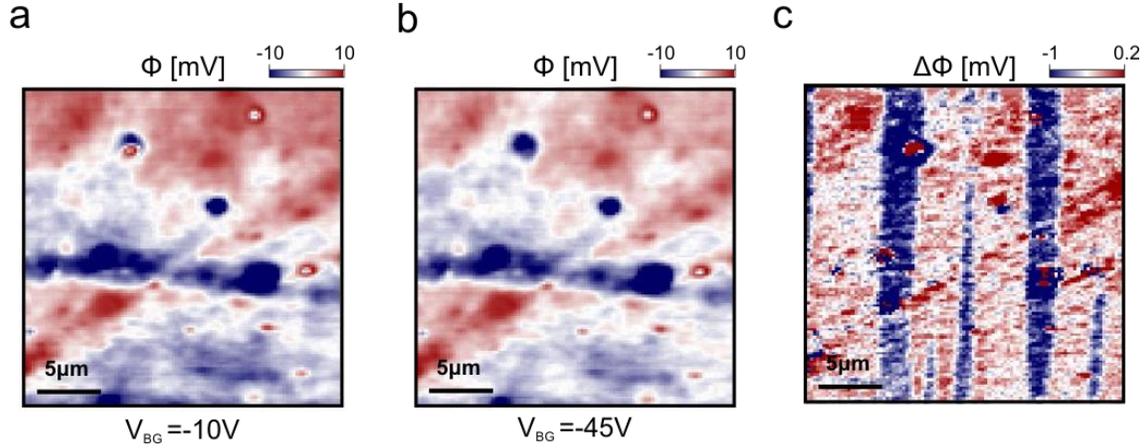

**Figure S5. Measured maps of the electrostatic disorder potential and their difference giving the domain-generated surface potential.** a-b) Electrostatic potential maps measured at $V_{BG} = -10V, -45V$. c) The difference between the data in panels (a) and (b). Since in panel (b) the entire field of view is occupied by a mono-domain, this difference map subtracts the disorder potential that is gate-voltage independent, allowing observation of the small potential modulations due to the striped domains present in panel (a) but hidden beneath the large disorder. The red $45^O$ streaks that are still visible in (c) are artifacts due to imperfect subtraction between (a) and (b). Such artifacts can arise from small, uncompensated gating differences of the SET by the back gate between images taken at such different gate voltages. It should also be stressed that in our studies we could not identify a clear influence of the point-like disorder features (like those observed *e.g.* in this figure) on the domain dynamics.

## S6. Optical images of domains: temperature and gate-voltage dependence

Figure S6a-b show measured optical images[2–5] in a bare STO sample measured at temperatures above the ferroelastic transition ($T = 110K$, Fig. S6a) and below the transition ($T = 70K$, Fig. S6b). The appearance of a striped pattern of domains similar in dimension and orientation to that observed in the scanning SET experiments is nicely

correlated with the crossing of this transition. We observe identical results in several LAO/STO and bare STO samples. Fig S6c shows a series of line cuts taken along the dashed orange line of S6b as a function of temperature. Red corresponds to higher optical intensity, whereas blue corresponds to lower optical intensity. For temperatures above the ferroelastic transition ($T = 105K$, marked by the black dashed line), there are no stripes in the image, indicating that the STO crystal is cubic. The stripes appear as the temperature drops below the transition, establishing that they arise from the STO ferroelastic transition to tetragonal crystal symmetry[3]. No other structural transitions are observed as the temperature is further lowered[6], though the lack of observation may be due to limitations of the optical setup.

Fig. S6d shows a diagram of the stripe mobility created from a collection of optical images of an LAO/STO sample. The mobility of the stripes is observed to be strongly temperature dependent. In the lower, gold region of the plot, the stripes are visible, but are immobile. As the gate voltage is made more negative and the temperature is lowered, the stripes become mobile, and we move into the "moving stripes" phase shown in green. As the gate voltage is made more negative still, or the temperature further lowered, the stripes retreat to the sample edges, and the center of the sample becomes homogenous with a mono-domain, shown in blue. This increasing domain wall mobility with decreasing temperature is consistent with the temperature dependence of the piezoelectric response measured by Grupp *et al.* (Reference 12 in main text).

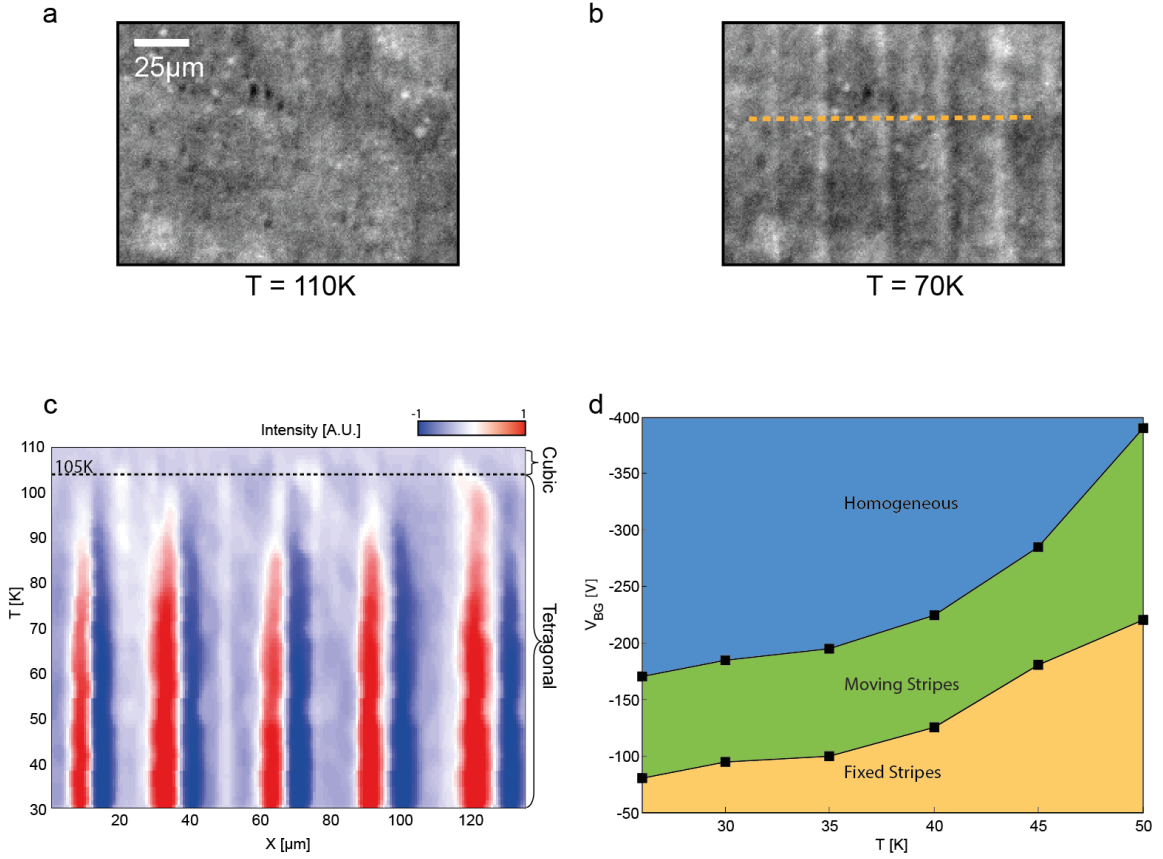

**Figure S6: Temperature dependence of stripes.** a,b) Optical images of bare STO. An image taken above the ferroelastic transition temperature (a) shows no stripes, while an image taken below the ferroelastic transition temperature (b) shows stripes[2–5]. c) Line cuts taken along the orange dashed line in (b) as a function of temperature. The stripes appear as the temperature drops below $T = 105K$, firmly establishing that they arise from the STO ferroelastic transition to tetragonal crystal symmetry[7]. d) A diagram of the stripe mobility created from a large set of optical images. The mobility of the stripes is observed to be strongly temperature dependent[8], increasing as the temperature is lowered and the gate voltage is made more negative. We find nice correspondence with the phase diagram measured in Ref 9 which used neutron diffraction.

## S7. Side view optical image of domains spanning entire sample thickness

Figure S7 presents a cross-section of bare STO imaged via cross-polarized optical microscopy at $T = 30K$ (below the ferroelastic transition temperature). We clearly see the X and Z domains separated by boundaries that penetrate the bulk of the STO at 45° angles (0° and 90° when viewed from above), many of which persist completely from top to bottom of the sample. Here we use the same domain labeling convention as explained in Fig. 3 of the main text.

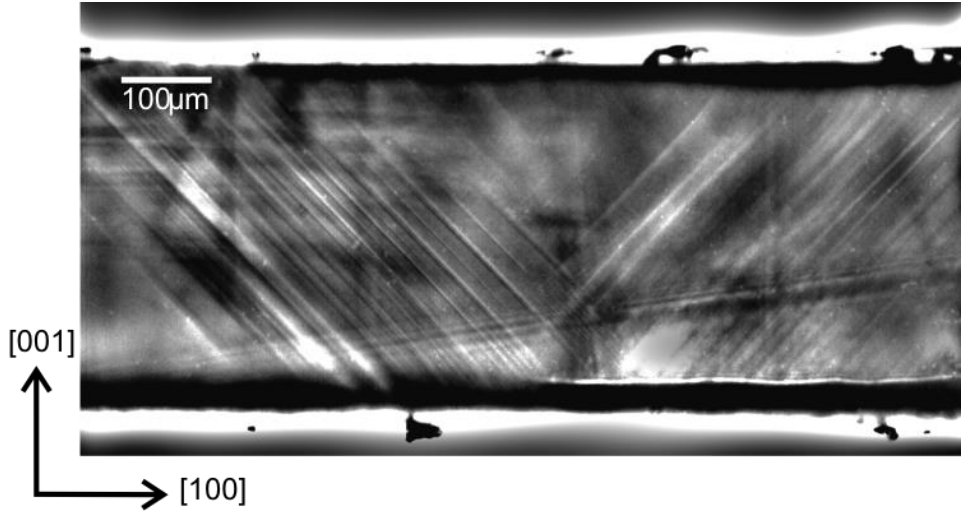

**Figure S7: Side view optical image of bare STO at T=30K.** The side plane of the sample is polished to allow viewing domain structure. Sample thickness is $500\mu m$. Bright top and bottom stripes in the figure correspond to the top and bottom planes of the sample. Clearly visible are 45° angled lines, which appear only below the ferroelastic transition temperature. On the contrary, the nearly vertical and horizontal lines are independent of temperature and correspond to scratches on the side surface.

**S8. The dependence of domain distribution on initial voltage cycling post cooldown**

In this section we show that the way in which the gate voltage is cycled after cooldown can be detrimental to the prevalence of striped domains in the sample. While as-cooled samples and those that are only cycled to small back-gate voltages remain covered with striped domains over their entire areas, if the back-gate voltage is cycled even once to a high enough voltage, the stripes can be cleared from the bulk of the sample, leaving only narrow sections of varying domains along the sample edges.

Upon cooling below the ferroelastic transition temperature, stripes appear distributed haphazardly over the entirety of the samples, emanating from their edges. After cycling the gate voltage, typically through hundreds of volts applied across the $\sim 500\mu m$ thick STO, we obtain repeatable gate-voltage dependent domain distributions[5]. Such gate voltage cycling can also clear the stripes entirely from the central areas of the samples and leave them only in a strip within a few hundreds of microns from the edges[10]. This phenomenology is summarized in Fig S8. Here, we cycle an LAO/STO sample at $T = 30K$ from zero volts on the back gate to negative gate voltages and then back to zero, and we repeat this procedure several times, each time increasing the magnitude of the negative voltage. Cycling to $V_{BG} = -50V$ and then back to $V_{BG} = 0V$ does not noticeably influence the domain distributions. However, increasing the cycling up to $V_{BG} = -100V$ clears some domains near the center of the sample (region enclosed by the yellow dashed line) even when the voltage is returned to $V_{BG} = 0V$. Cycling up to $V_{BG} = -200V$ and back to $V_{BG} = 0V$ leaves the sample free of domain walls except for strips within $\sim 100 - 200 \mu m$ from the edges (more generally can be within $100 - 500 \mu m$ from the edges depending on sample-specific details).

Although we observed identical domain behavior in all measured samples, we note that the domain dynamics are temperature dependent (Supplementary S6), should depend on built-in strain and specific sample growth conditions, and also may be different near patterned features. These aspects should be carefully tested in future transport experiments carried out in conjunction with direct observation of domain dynamics.

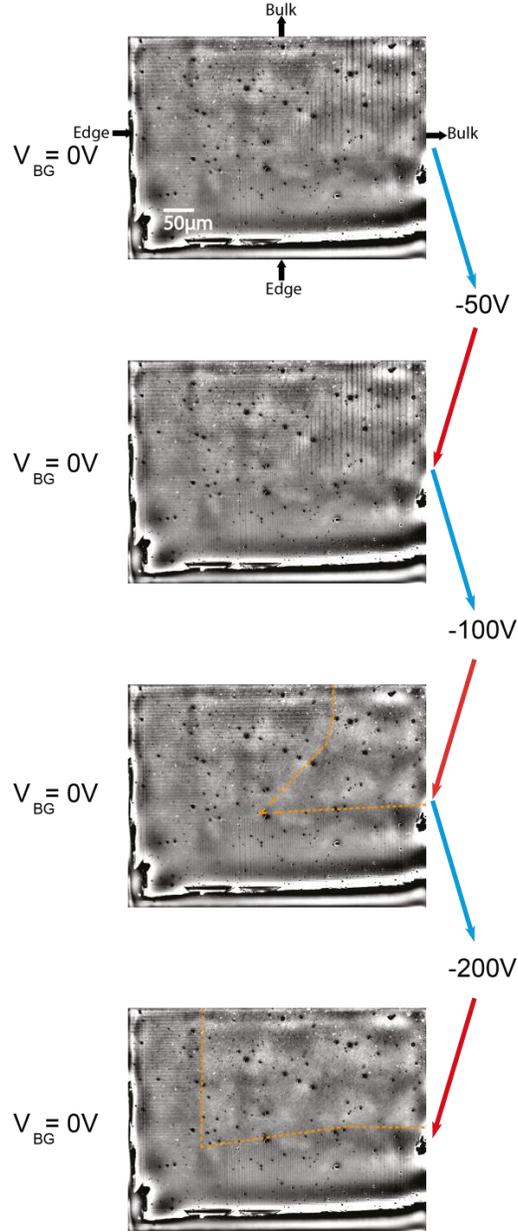

**Figure S8: The dependence of domain distribution on initial voltage cycling post cooldown.** Optical images of an LAO/STO sample at $T = 30K$ as it is cycled from zero gate voltage post-cooldown to negative voltages and back to zero, increasing the magnitude of the negative voltage with each iteration (arrows on the right). Before voltage cycling, stripes are distributed haphazardly throughout the sample (top panel). Cycling the gate voltage to $V_{BG} = -50V$ and back to $V_{BG} = 0V$ does not noticeably influence the domain distribution (2$^{nd}$ panel from top). Cycling the gate voltage up to $V_{BG} = -100V$ and back to $V_{BG} = 0V$ begins to clear stripes from the bulk of the sample (region bounded by dashed orange line, 2$^{nd}$ panel from bottom), and cycling up to $V_{BG} = -200V$ and back to $V_{BG} = 0V$ clears stripes everywhere except for near the edges (bottom panel).

## S9. Comparison between samples from different suppliers and with different surface treatments

To investigate how general are our observations regarding the voltage dependence of the domain structure, we performed optical measurements across a variety of different samples. In addition to two different LAO/STO samples, we also studied several bare STO(100) samples obtained from different substrate suppliers (Crystec, MTI) which underwent different surface treatments (untreated vs. $TiO_3$-terminated surfaces created via step-etching and thermal annealing).

Across all samples, we observed a rather universal dependence of the domain landscape upon gate voltage cycling: the as-cooled samples (before the application of any gate voltage) have their entire area densely covered by differently-oriented domains that appear at the ferroelastic transition temperature and maintain the same landscape when the sample is cooled further to lower temperatures. Upon cycling the relative voltage between the back gate and the surface to high values and then back to zero, we observe that in all cases 0° and 90° domain wall patterns (between x/z and y/z domains, respectively) retreat toward the edges of the samples, leaving the centers free of these domain walls, even when the gate voltage is returned to zero. This behavior is demonstrated for three different samples (LAO/STO, $TiO_3$-terminated STO from Crystec, untreated STO from MTI) in Figure S9 below. The left panels show the pattern of domains before voltage cycling (as-cooled), and as can clearly be seen the domain wall patterns cover the entire field of view in the figure (in all cases they cover the entire area of the sample, not shown). The right panels show the domain landscape after cycling to $V_{BG} = +200V$ and back to $V_{BG} = 0V$. In all cases the 0° and 90° domain walls shrink to be confined within only a short distance from the sample edge, as indicated by the orange lines which mark the boundary where these patterns terminate. Note that the additional (voltage-cycling-independent) structure in the sample in panel S9b is due to surface contamination, and that in the MTI sample (untreated, bare STO(100)) in panel S9c, the signal from domain walls between x/y domains (45°) that originates deep below the surface is very strong and remains dominant when the focus is on the surface. Small patches of x/z domain walls still remain in this sample after voltage cycling.

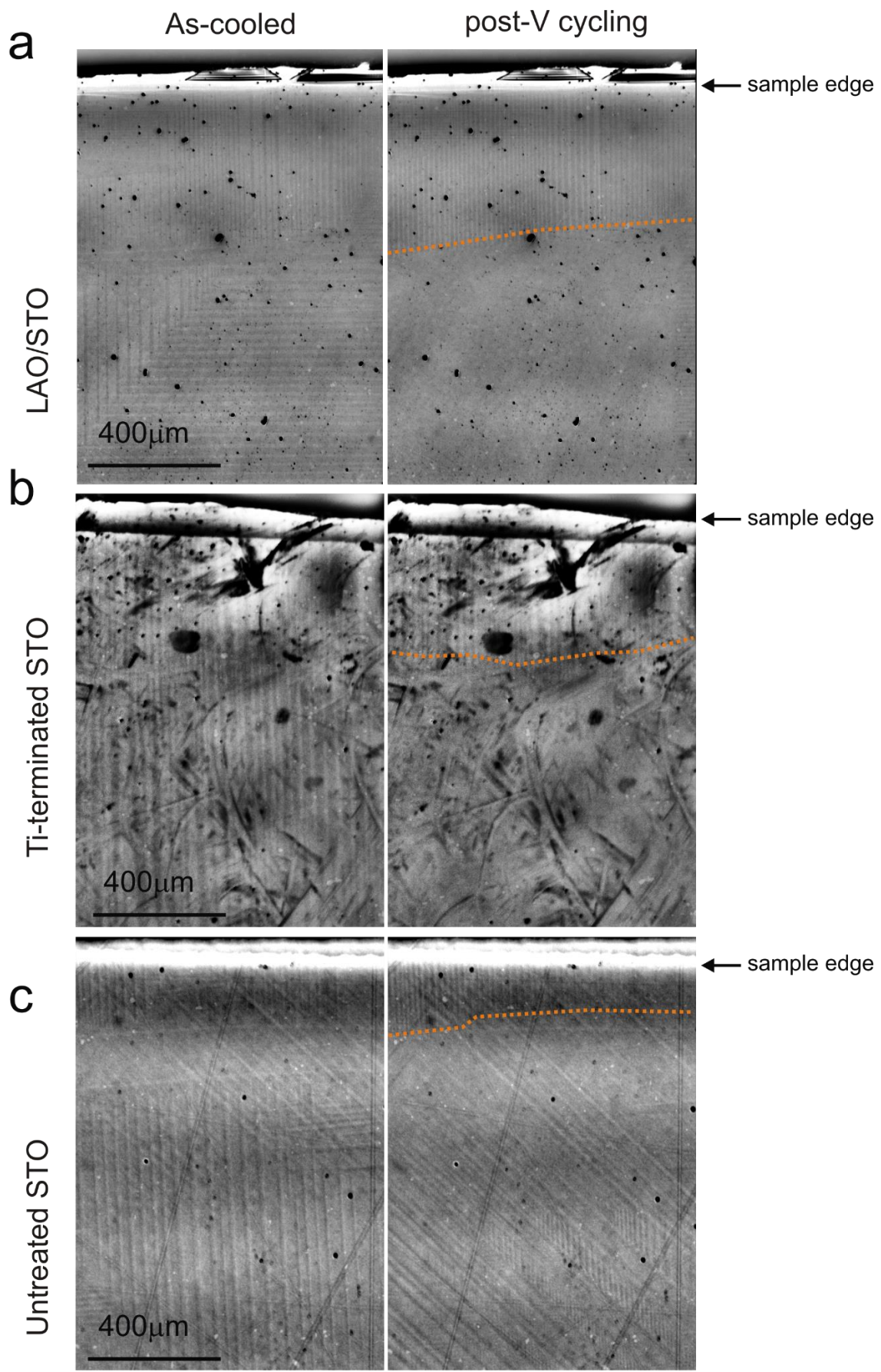

**Figure S9: Images of domains as-cooled and post-voltage cycling in several different samples.** Each panel presents a different sample, before (left) and after (right) voltage cycling. a) 5 unit cell LAO/STO sample. b) TiO$_3$-terminated STO(100) from Crystec. The titanium surface termination is achieved through an etch process, followed by thermal annealing. c) Untreated, as-grown STO(100) from MTI Corporation. In all panels the edge of the sample appears at the top of the image (black arrow). The dashed orange lines mark the boundaries of the regions cleared of x/z and y/z domain walls after voltage cycling of the back gate. The spatial scale is identical in all images. The STO thickness in all cases is 500$\mu m$. The visible smudges in panel (b) are due to dirt that condensed on the sample surface during one of the cooldowns, which is temperature and gate-voltage independent, and unrelated to the underlying domains. In the untreated STO sample in panel (c), 45° domain boundaries between x/y domains gave an especially strong signal as compared with the other two samples. These 45° domain boundaries are usually seen when the focal plane is deep below the sample surface, whereas at the surface only the 0° and 90° x/z and y/z domain walls are typically visible. In this sample the signal of the deeper 45° domain boundaries was so strong that it had significant visual imprint even when the focal plane was at the surface of the sample.

## S10. Transport and capacitance characterization of the 2DES of the studied LAO/STO sample

To support the notion that our sample is typical of those used in other studies, we have further characterized the LAO/STO sample on which the scanning SET measurements were made by performing a series of transport and capacitance measurements. Magnetotransport measurements were performed utilizing the Van der Pauw method to extract the mobility and carrier density of the 2DES, and the capacitance between the 2DES and the back gate was also measured as a function of gate voltage.

After voltage cycling, at $V_{BG} = +200V$ the carrier density of the 2DES was $2.2 \pm 0.1 \times 10^{13} cm^{-2}$, with mobility $660 \pm 10 \ cm^2/Vs$. The sheet resistance as a function of gate voltage is shown in figure S10a. At the highest 2DES density ($V_{BG} = +200V$), this sheet resistance is lowest at a value of 400 $\Omega/\square$. As the gate voltage is made more negative and the 2DES is depleted, the sheet resistance increases to the value of 6000 $\Omega/\square$ at $V_{BG} = -200V$.

The capacitance between the 2DES and the back gate is plotted as a function of voltage in figure S10b. The capacitance is peaked around zero gate voltage where the dielectric constant of the STO is highest. Because the STO dielectric constant decreases with increasing applied field, the capacitance drops as the gate voltage is swept away from zero in either direction. We note that the peak in capacitance is always located near zero voltage, independent of voltage cycling history. This suggests that oxygen migration plays a minimal role in our cryogenic measurements. Were oxygen migration to occur as a result of voltage cycling, we would expect the resulting oxygen vacancies to act as an effective gate voltage, shifting the peak in capacitance away from zero volts, which we do not see in our measurements.

Overall, these trends in transport and capacitance compare favorably to such measurements in other LAO/STO studies, supporting the relevance of the domain physics uncovered in this work to other 2DES systems built upon STO.

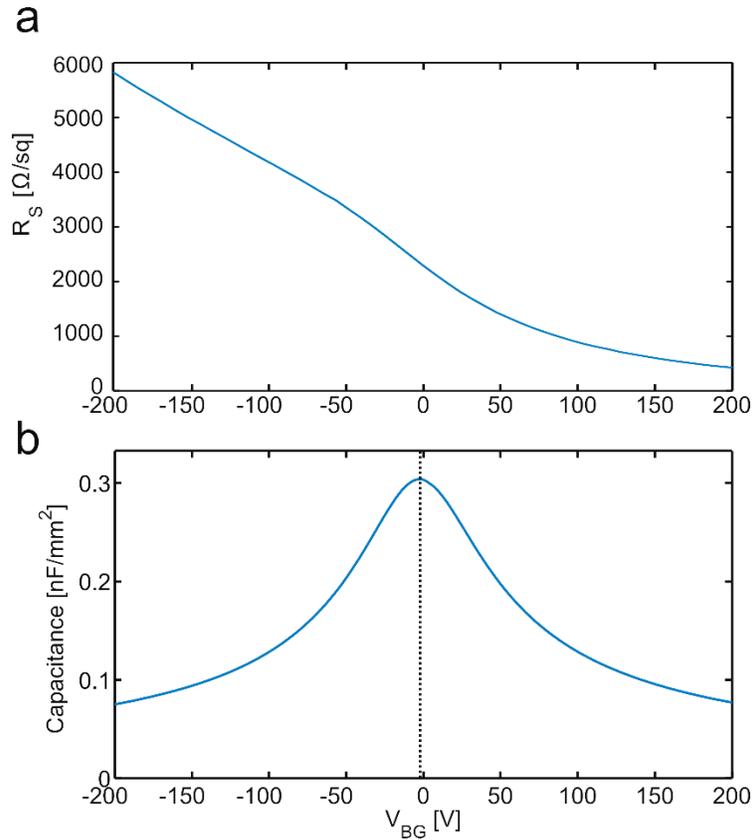

**Figure S10: LAO/STO 2DES characterization: sheet resistance and capacitance.** a) Sheet resistance of the 2DES as a function of back gate voltage after voltage cycling for the 6 unit cell LAO/STO sample used for SET imaging in the main text. The resistance steadily increases as the gate voltage is made more negative, depleting the 2DES electron density. b) Capacitance of the same sample as a function of back gate voltage post voltage cycling. The peak capacitance is marked by a dashed vertical line.

### S11. Surface potential and 2DES density modulation

In this section, we explain the connection between the local surface potential variation measured by the SET and the corresponding density modulation in the 2DES. In principle, the surface of two neighboring but separated domains (depicted schematically as $A$ and $B$ in Fig. S11a) may have different electrochemical potentials, *e.g.* due to changes of the 2DES energy bands with mechanical stretching along different directions. If these two domains are in electrochemical equilibrium (their natural state when in contact, Fig S11b) they must have the same electrochemical potential. This equilibrium is achieved by charge transferring across the domain wall, forming a dipole-like charge distribution at the wall that generates a step in the local vacuum level. This potential step exactly compensates the workfunction difference, $\Delta \phi = W_A - W_B$, thereby flattening the electrochemical potential across the domains (Fig S11b). This step in the local vacuum potential across domain walls is the quantity measured in our scanning SET experiment.

The amount of density modulation in the dipole-like charge double layer in the 2DES is estimated from simple electrostatics. As this charge redistribution is restricted to 2D whereas the field lines propagate from one domain to its neighbor in 3D, the charge density will decay as one over the distance from the domain wall: $\Delta n \sim \epsilon \Delta \phi / e d$, where $\epsilon$ is the dielectric constant, $e$ is the electron charge, and $d$ is the distance from the domain wall (Fig S11c). Thus for lengthscales on the order of typical ferroelastic domain wall thickness $\sim 1 - 10 nm$ and an $\epsilon \sim 1000 \cdot \epsilon_0$, which would correspond to the reduced dielectric constant of STO in the high field region near the interface, we find $\Delta n \sim 0.5 - 5 \cdot 10^{12} cm^{-2}$. This would be a significant fraction of the average charge density in the 2DES.

We note that above we consider only the simplest and most generic mechanism for charge modulation in the system induced by the potential modulation. Other, more speculative scenarios that include strain effects on the LAO and charge redistribution between internal degrees of freedom in the system (e.g. between different subbands of the 2DES) may also be at play. Such mechanisms could contribute additional channels for creating significant modulation to the 2DES density in excess of the mechanism discussed above.

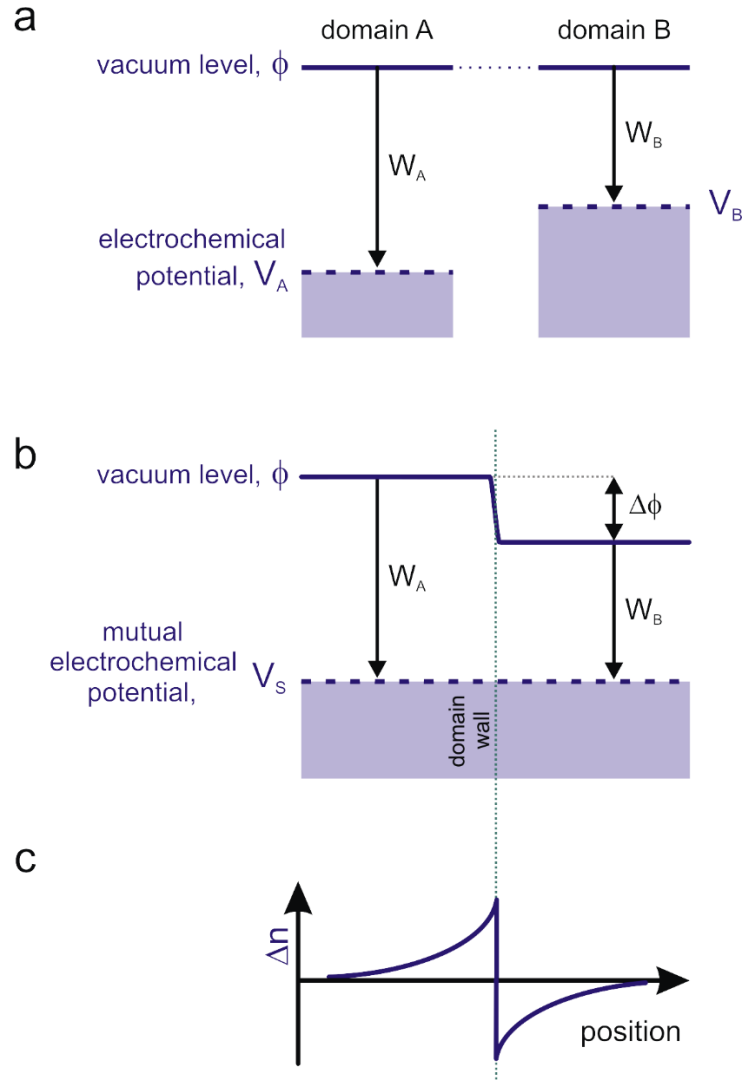

**Figure S11: Potential and charge modulation across a domain wall.** a) Separated domains $A$ and $B$ may have 2DES with different electrochemical potentials $V_A$ and $V_B$, respectively, due to strain-induced changes to the 2DES bandstructure, resulting in differing workfunctions $W_A$ and $W_B$. b) When in contact, the domains reach equilibrium by transferring charge across the domain wall to maintain a constant value of the electrochemical potential, $V_S$. This generates a step (contact potential) in the vacuum potential, $\phi$,

equal to the difference in workfunctions: $\Delta \phi = W_A - W_B$. c) The corresponding charge density modulation that creates this contact potential is dipole-like, falling off as one over the distance from the domain wall, since the charge redistribution is confined to 2D whereas the field lines between domains permeate the entire 3D volume.